# Chimeric protein complexes in hybrid species generate novel evolutionary phenotypes.

Elzbieta M. Piatkowska[1], David Knight[1] and Daniela Delneri[1*]

[1]Faculty of Life Sciences, Michael Smith Building, University of Manchester, United Kingdom.

[*]Corresponding author, e-mail: d.delneri@manchester.ac.uk

Running title: Proteome cross-talk impact on hybrid fitness.

**Abstract**

Hybridization between species is an important mechanism for the origin of novel lineages and adaptation to new environments. Increased allelic variation and modification of the transcriptional network are the two recognized forces currently deemed to be responsible for the phenotypic properties seen in hybrids. However, since the majority of the biological functions in a cell are carried out by protein complexes, inter-specific protein assemblies therefore represent another important source of natural variation upon which evolutionary forces can act. Here we studied the composition of six protein complexes in two different *Saccharomyces* "sensu strictu" hybrids, to understand whether chimeric interactions can be freely formed in the cell in spite of species-specific co-evolutionary forces, and whether the different types of complexes cause a change in hybrid fitness.

The protein assemblies were isolated from the hybrids via affinity chromatography and identified via mass spectrometry. For the first time, we found evidence of spontaneous chimericity for four of the six protein assemblies tested and we showed that different types of complexes can cause a variety of phenotypes in selected environments. In the case of *TRP2/TRP3* complex, the effect of such chimeric formation resulted in the fitness advantage of the hybrid in an environment lacking tryptophan, while only one type of parental combination of the MBF complex could confer viability to the hybrid under respiratory conditions. These phenotypes were dependent on both genetic and environmental backgrounds.

This study provides the first empirical evidence that chimeric protein complexes can freely assemble in cells and reveals a new mechanism to generate phenotypic novelty and plasticity in hybrids to complement the genomic innovation resulting from gene duplication. The ability to exchange orthologous members has also important implications for the adaptation and subsequent genome evolution of the hybrids in terms of pattern of gene loss.

**Author summary**

The *Saccharomyces cerevisiae* "sensu stricto" group represent an excellent example of sister species which can readily hybridise to occupy new ecological niches. Hybrids harbour the DNA of both parents and can display diverse pattern of gene expression. Less is known about the protein interactions that occur in hybrids, where two diverged proteome co-exist and are responsible for the correct execution of the biological function. In fact, hybrids could potentially form different chimeric variants of the same protein complex by using all the different combinations of parental alleles available. Chimeric interactions are expected to be sub-optimal and therefore discouraged since the members forming the protein complex are from different parents and have a different evolutionary history. Interestingly, here, we show experimentally that chimeric protein assemblies are spontaneously established in different yeast hybrids, and that such chimericity produces different phenotypic variants displaying loss or gain of fitness according to their genetic background and to the environment that they are exposed. These findings imply that the formation of chimeric complexes offers a new source of natural variation, widens the adaptation potential of the hybrids towards new nutritional contexts, and may influence genome evolution through selective retention of optimal alleles.

**INTRODUCTION**

The *Saccharomyces sensu stricto* yeasts represent a diverse, monophyletic group of species that have the ability to produce viable and stable hybrids that can propagate mitotically. Hybrids among yeast species and strains seem to be very common, especially amongst wine, and beer brewing yeasts[1,2], but also within natural ecological niches[3]. When two parental genomes merge in yeast hybrids there is a potential for genetic novelty but also for a genetic conflict to occur. Dominant genetic incompatibilities do not seem to occur in the *S. cerevisiae sensu stricto* group[4], however evidence of recessive allelic incompatibilities between nuclear and mitochondrial genomes have recently been uncovered[5].

Hybridisation can play an important role in evolution since hybrids could occupy a different niche from both parental species and eventually establish a new lineage. The presence of naturally occurring yeast hybrids isolated from specific environments seem to confirms this hypothesis[6,7]. So far, many unique characteristics of the *Saccharomyces* "sensu stricto" species and hybrids have been attributed to changes in gene expression, including novel *cis-trans* interactions[8] and to divergence in regulatory regions[9]. Nevertheless, in the hybrid cellular environment, where two sets of homologous proteomes coexist, there is also the potential for the cell to form chimeric assemblies between homologus protein complexes. Analysis of large-scale proteomics data has shown that the majority of cellular processes are carried out by protein assemblies rather than single proteins and that over 60% of yeast proteins form obligate complexes[10]. Since the correct formation of a complex is essential to carry out the biological function, we would expect that any sub-optimal protein interaction would be detrimental to the cell and therefore discouraged by the cell. On the other hand, spontaneous chimeric assemblies may widen the adaptation potential of the cell, since several different combinations of the same protein complex can be used. Therefore, such situation can lead to new phenotypic variants that are beneficial to the hybrid in novel contexts.

The primary aim of this work is to establish proof of principle that chimeric protein complexes can form freely in hybrids of *Saccharomyces* species despite the intra-specific co-evolutionary forces and to quantify the impact that such complexes can have on the overall fitness of the hybrids. In fact, chimericity in protein-protein interaction represents a potentially important mechanism for generating phenotypic diversity upon which evolutionary forces can act, and may constitute a molecular explanation of hybrid vigour.

**RESULTS AND DISCUSSION**

**Experimental strategy for the analysis of chimeric complexes in yeast hybrids**

To test for the existence of natural chimeric complexes in yeast hybrids, we analysed six physically stable 'obligatory' protein complexes (Table S1 of supplementary materials) each of which have constitutively expressed members that were previously recovered by large-scale protein interaction studies and also by independent small-scale biochemical studies[11,12].

We created *S cerevisiae/S. mikatae* (*Sc/Sm*) and *S. cerevisiae/S. bayanus* (*Sc/Sb*) hybrids by crossing either *S. mikatae* or *S. bayanus* with *S. cerevisiae* strains carrying a molecular tag (TAP-tag) at the C-terminus of a selected member of the protein complex (see Fig. S1 of the supplementary materials). Tagged proteins, along with their interacting partners, were isolated via affinity chromatography and all the members of the protein complex were identified via mass spectrometry. If only species-specific parental complexes are established in the hybrid, just proteins from the species carrying the TAP-tag (*S. cerevisiae*) will be identified. However, if chimeric protein complexes are formed, proteins from the other parental species (*S. mikatae* or *S. bayanus*) will also be isolated and identified (Fig. 1). The protein fractions were analyzed by mass spectrometry to identify tryptic peptides in a custom protein database of six *Saccharomyces sensu stricto* yeast proteomes. Species-specific peptides were distinguished from the shared peptides that are identical between the two parental species. As control experiment to test whether *in vitro* chimeric interactions were generated artefactually during the protein extraction procedure (as opposed to *in vivo* within the hybrid cellular environment), a mixture of parental cells (i.e. *S. cerevisiae* and *S. mikatae* or *S. bayanus*) were grown separately and mixed together just prior to cell lysis. To establish that both parental genomes were present, all hybrids were screened for chromosomal content via PCR using species-specific primers (see Fig. S2 of the supplementary materials). Transcription of the homologous members of the protein complexes in the hybrids was also confirmed via RT-PCR (see Figures S3-S8 of the supplementary materials).

**Analysis of the nature of the protein complexes in yeast hybrids**

The first complex we considered was the Sec 62/63 complex, a tetramer that is involved in the transport of proteins across the ER membrane, composed of two essential proteins, Sec62p and Sec63p and two non-essential proteins, Sec66p and Sec72p[13]. In both hybrids *Sc/Sm* and *Sc/Sb*, the mass spectrometry analysis identified Sec63p and Sec72p from either *S. mikatae* or *S. bayanus,* respectively, demonstrating that in yeast hybrids the assembly of the Sec62/63 complex can be spontaneously chimeric, (Fig. 3, and Fig. S9-S16 supplementary materials, Table S2 and S3).

Evidence of chimeric interactions were also detected between members of the *TRP2/TRP3* complex, involved in the tryptophan biosynthesis[14] (Figures S17 and S18, Tables S4 and S5) and the CTK complex, involved in transcription and translation regulation[15] (Ctk1p, Ctk2p, Ctk3p; see Figure S19 and S20, Table S6 and S7), in both *Sc/Sm* and *Sc/Sb* hybrids.

In the case of the MBF complex, a dimer composed of two proteins, Mbp1 (a transcription factor responsible for DNA synthesis at the G1/S phase of the cell cycle) and Swi6 (a *trans*-activating component)[16], chimeric complexes were only identified in hybrids *Sc/Sb*, while, surprisingly, no free interaction was detected in the hybrids of the more closely related species *S. cerevisiae* and *S. mikatae* (Figure S21 and S22, Tables S8 and S9). This results indicates that, given the choice, Mbp1p from *Sc* prefer to form uni-specific complexes with Swi6p from *Sc* in *Sc/Sm* background. When considering protein-protein interactions the sequence identity of the biding interfaces is likely to be more important than the phylogenetic relationship. In fact, Swi6p shows greater sequence similarity between *S. cerevisiae* and *S. bayanus* than between *S. cerevisiae* and *S. mikatae*, despite their phylogeny (data not shown).

The remaining two complexes tested, the RAM (Ram1p and Ram2p, farnesyltransferase complex involved in the prenylation of Ras proteins)[17] and KU (Yku70p and Yku80p), involved in double strand breaks repair and non-homologous end joining)[18], appeared unable to form chimeric complexes in any hybrid background. In fact, using Yku70p as TAP-bait, no specific Yku80p peptides from *S. bayanus* and *S. mikatae* parental species were ever found in any biological replica tested, while numerous *S. cerevisiae* specific Yku80p peptides were consistently isolated (Tables S10-S13). Although the failure to detect such interactions in mass spectrometry is not a definite proof that chimeric complexes are not at all assembled, this data suggests that chimericity within RAM and Ku complexes may at least occur rarely, and that the proteins forming such complexes tend to assemble in uni-specific manner if given the option. Interestingly, an independent study of the KU complex in hybrids of two diverged strains of *S. paradoxus* showed that negative epistatic interactions occur between the different homologues of Yku70p and Yku80p, suggesting either lack of assembly or functionality of the heterodimer[19]. The inability to detect spontaneous chimeric complex formation in both *Sc/Sm* and *Sc/Sb* hybrids observed in this work support the idea that the prevention of complex formation could be the possible mechanism for the negative epistasis identified between Yku70p and Yku80p in the *S. paradoxus* strains.

**Phenotypic variations caused by different types of protein assemblies**

We evaluated the impact that chimeric interactions have on fitness by forcing the hybrids to use only one specific type of complex to carry out the biological function. We chose to investigate the *TRP2/TRP3* ad the MBF complex, since the relationship between the functional complexes and the resulting output fitness could be clearly measured under tryptophan starvation and respiratory growth condition, respectively. In fact, the *TRP2/TRP3* complex is involved in the first step of the tryptophan biosynthesis[14], and null mutants of Mbp1p and Swi6p display a range of fitness defects including decrease rate or respiratory growth and abnormal mitochondrial morphology[20].

We created different combinations of the *TRP2/TRP3* and MBF complexes by deleting different protein members in both *Sc/Sm* and *Sc/Sb* hybrid backgrounds (Figure 3A and 4A), and then scored the growth rates of the hybrids carrying either uni-specific or chimeric complexes.

For the *TRP2/TRP3* complex in the *Sc/Sb* background, a large range of fitness levels was detected for the different types of assemblies, and in particular the chimeric complex of Trp2p$^{Sb}$/Trp3p$^{Sc}$ grew much better than the uni-parental hemizygous controls Trp2p$^{Sb}$/Trp3p$^{Sb}$ and Trp2p$^{Sc}$/Trp3p$^{Sc}$ in a medium lacking tryptophan (Figure 3B). Strains containing the alternative hybrid complex Trp2p$^{Sc}$/Trp3p$^{Sb}$ grew as well as the *S. cerevisae* Trp2p$^{Sc}$/Trp3p$^{Sc}$ hemizygous control and better than the *S. bayanus* Trp2p$^{Sb}$/Trp3p$^{Sb}$ strain, while the original *Sc/Sb* hybrid strain showed an intermediate fitness (Figure 3B). To confirm the increased fitness of the strain expressing a Trp2p$^{Sb}$/Trp3p$^{Sc}$ chimeric complex, competition experiments between the chimeric hybrids and a GFP reference strain was carried out using FACS analysis[21]. The results showed that strains with the chimeric Trp2p$^{Sb}$/Trp3p$^{Sc}$ complex were more fit than those with the other chimeric complex (Trp2p$^{Sc}$/Trp3p$^{Sb}$) and those with both uni-specific protein-protein interaction combinations (Figure 3C).

For the MBF complex in the *Sc/Sb* background only the hybrid carrying the uni-specific combination Mbp1p$^{Sb}$ and Swi6p$^{Sb}$ derived from *S. bayanus* was able to grow in media containing glycerol, a carbon source that can only be respired (Figure 4). The other inviable parental combination of Mbp1p$^{Sc}$/Swi6p$^{Sc}$ could not be rescued by adding either Mbp1p$^{Sb}$ or Swi6p$^{Sb}$ to its genotype, showing that the presence of both *S. bayanus* members of the MBF complex is required for hybrid viability (Fig. S23). Interestingly, the restriction analysis of the mitochondrial genes *COX2* and *COX3* indicated that the *Sc/Sb* hybrids harbour the *Sb* mitochondrial DNA (data not shown). This example not only shows phenotypic plasticity of different chimeric assemblies, but also represents a novel case of hybrid incompatibility between *S. cerevisiae* and *S. bayanus*.

Fitness variation between the different types of protein assemblies was not otherwise observed in *Sc/Sm* hybrids either for the *TRP2/TRP3* or for the MBF complex (Figure S24), underlying the background dependency of these phenotypes.

**Conclusions**

Here we have shown that protein complexes in yeast hybrids are able to spontaneously exchange components for inter-specific orthologs. Out of the six complexes studied four were convincingly found to form natural chimeric protein assemblies in either one or both genetic hybrid background (i.e. Sec62-63,

*TRP2/TRP3*, *MBF,* and CTK complex). These results provide the first proof of principle that chimeric protein interactions in hybrids can arise to generate evolutionary novelty in protein-protein interaction networks, providing a new evolutionary mechanism to complement innovation by gene duplication[22].

We also found that some complexes prefer to form species-specific configurations in the natural hybrid cell environment (i.e. Ku and RAM complex). The lack of spontaneous chimeric assembly in these cases could be due to less favourable changes in the binding interfaces of the proteins, or to stoichiometry imbalance between homologous proteins in the hybrid[23]. The inability to create chimeric interaction can be responsible for some negative epistatic effect seen in hybrids[19].

We showed that different type of complexes can cause a variety of phenotypes in selected environments. In the case of *TRP2/TRP3*, we find that chimeric complex formation can lead to hybrid vigour, reinforcing the idea that the ability to form different types of protein assemblies could be advantageous in specific nutritional contexts. In the case of MBF complex only one parental combination of protein-protein interaction was compatible with cell viability under respiratory condition, highlighting a new case of allelic incompatibilities in yeast hybrids. These phenotypes were proved to be dependent on both genetic and environmental backgrounds since we did not observe any fitness change in *Sc/Sm* hybrids and the advantages could be lost or gained in different media, such as in the case of the strains carrying different combination of the MBF complex grown in YPD or YP-glycerol (Fig. 4B).

Ultimately, this study proposes a novel molecular mechanism for creating phenotypic variation within a hybrid cell, with important implications for understanding the evolutionary forces that govern the reshaping of hybrid genomes.

The genomic fate of the homolog genes will in fact be influenced by the ability or not of the hybrid to create inter-specific protein assemblies (Fig. S25). Moreover, chimeric complexes may be able to recruit new proteins and evolve new functions in the cell[24]. In the future, the genomic information of naturally occurring hybrids (like *S. pastorianus* strains) will provide insight into the nature of how the formation of chimeric interactions influences selective gene retention of members of protein complexes and networks.

**MATERIALS AND METHODS**

**Generation of yeast hybrids**

All the TAP-tagged strains were obtained from the EUROSCARF strains collection (http://web.uni-frankfurt.de/fb15/mikro/euroscarf/cellzome.html). Hybrids between *S. cerevisiae* strains (bearing the TAP-tag in selected members of different protein complexes) and wild-type *S. mikatae* 1815 and *S. bayanus* NCYC2669 species were generated using a Singer Instruments MSM micromanipulator as previously described[25]. To enable selection of hybrid colonies, we made the *S. cerevisiae* TAP strains geneticin-resistant by inserting a kanMX in the neutral *AAD3* locus. Hybrid colonies were then selected on minimal media containing geneticin G418 (see Figure S1). The nature of the chromosomes were verified by chromosomal PCR using genomic DNA from the hybrid as template and species-specific primers designed to distinguish between *S. cerevisiae*, *S. mikatae* and *S. bayanus* alleles (see Figure S2, primer sequences available on request). Hybrid genomic DNA and RNA was isolated using the DNasy Blood & Tissue kit and the RNeasy mini kit (Qiagen, Crawley, UK), respectively.

**Purification of protein complexes from yeast hybrids and mass spectrometry analysis**

Purification of the protein complexes was carried out using the standard TAP protocol[26] optimized for these specific classes of proteins. In particular, two affinity binding steps, the IgG Sepharose and Calmodulin Binding Protein (CBP) binding and TEV protease cleavage were carried out for 2 hours at 4 $^o$C instead of 16 $^o$C. The protein mixtures were resolved using 1D gel electrophoresis, stained with Coomassie Bio Safe (Bio-Rad) and digested with trypsin (Promega). The trypsin digest was carried out overnight at 37 $^o$C according to Shevchenko, A. *et al.* (ref. 27). The digested protein mixture was separated by the high performance liquid chromatography (HPLC) and analyzed by tandem mass spectrometry (ESI MS/MS) (Micromass CapLC-Q-ToF, Waters, Manchester, UK). Spectra acquired for every protein complex member were compared against a custom database containing all proteins from *S. cerevisiae* "sensu stricto" species, using Mascot version 2.2.06 (Matrix Science Inc., Boston, MA). Scaffold (Scaffold_2_01_00, Proteome Software Inc., Portland, OR) was used to validate MS/MS based peptide identification. A peptide match was acknowledged if it could be established at greater than 50.0% probability as specified by the Peptide Prophet algorithm[28]. Protein identifications were accepted if they could be established at greater than 95.0% probability by Protein Prophet and contained at least 2 identified peptides. The Liverpool Peptide Mapping Tool (http://www.liv.ac.uk/pfg/Tools/Pmap/pmap.html) was used to generate proteolytic peptide maps of protein complex members. The peptide maps were generated with one trypsin miscleavage per site after lysine and arginine (K-X, R-X) but not at lysine-proline and arginine-proline (K-P, R-P) sites.

**Generation of chimeric protein complexes in *Sc/Sm* and *Sc/Sb* hybrids and fitness assays.**

Chimeric and unispecific versions of the *TRP2/TRP3* and MBX complexes in both *Sc/Sm* and *Sc/Sb* hybrids were generated by PCR-mediated gene deletion strategy using hygromycin (*HPH*) and nourseothricin (*NAT*) as selectable markers[29]. The *S. cerevisiae TRP2* and *TRP3* copies were replaced with *HPH* while the *S. bayanus* ones were deleted using *NAT* (see Figure 3). Similarly for the MBF complex, the *S. cerevisiae* homologs of Mbp1 and Swi6 were disrupted using *HPH*, while the *S. bayanus* copies of Mbp1 and Swi6 were deleted using *NAT* (see Figure 4). All the primers used to create the hybrid strains carrying different versions of protein complexes are available on request.

Yeast hybrids were grown in YPD and minimal F1 media[30] at 30 °C for 40 hours with continuous shaking. Growth rates were measured by absorbance at $OD_{595}$ at 5 minutes intervals using Fluostar Optima bioscreen workstation (BMG Labtech).

Fitness competition assays were carried out by FACS analysis according to Lang *et al.* (ref. 21). As reference strain we used the FY3 strains bearing the GFP tag at the C-terminus of CDC33p (generated for the purpose of this experiment), and the competition was carried out in F1 media lacking tryptophan. The hybrids strains were mixed with the reference strain in 4:1 ratio, and a total of $1 \times 10^5$ cells, counted on a cellometer (Auto M10, Nexcelom), were inoculated into a 1 ml of fresh medium. The strains were allowed to grow for 12 hours and then the ratio of the number of hybrid cells over the fluorescent reference was determined using the Dako CyAn flow cytometer, with a total counting total 50,000 cells for each time point. Three biological and three technical replicates were performed for each fitness measurement. The $s_g$ fitness coefficient was calculated using the following equation:

$$S_g = \frac{\ln(H_f/R_f) - \ln(H_0/R_0)}{g_f - g_0}$$

where, H and R are the cell number of the hybrid and reference strain and $g_0$ and $g_f$ are the number of generations at the beginning and after a time interval (12 hours).

**ACKNOWLEDGMENTS**

The authors wish to thank Andrew W. Murray and John Koschwanez for advice and expert assistance with the FACS competition assay; Julian Selley for constructing the custom Mascot library; Casey M. Bergman, Simon J. Hubbard, and Catherine Millar for the critical reading of the manuscript.

**REFERENCES**

1. Masneuf I, Hansen J, Groth C, Piskur J, Dubourdieu D (1998) New hybrids between *Saccharomyces* sensu stricto yeast species found among wine and cider production strains. Appl Environ Microbiol 64: 3887-3892.

2. Gangl H, Batusic M, Tscheik G, Tiefenbrunner W, Hack C, *et al.* (2009) Exceptional fermentation characteristics of natural hybrids from *Saccharomyces cerevisiae* and *S. kudriavzevii*. Nat Biotechnol. 25: 244-251.

3. Liti G, Carter DM, Moses AM, Warringer J, Parts L, *et al.* (2009) Population genomics of domestic and wild yeasts. Nature 458: 337-341.

4. Greig D, Borts RH, Louis EJ, Travisano M. (2002) Epistasis and hybrid sterility in *Saccharomyces*. Proc Biol Sci 269: 1167-1171.

5. Lee HY, Chou JY, Cheong L, Chang NH, Yang SY, et al. (2008) Incompatibility of nuclear and mitochondrial genomes causes hybrid sterility between two yeast species. Cell 135: 1065-1073.

6. Blieck L, Toye G, Dumortier F, Verstrepen KJ, Delvaux FR, et al. (2007) Isolation and characterization of brewer's yeast variants with improved fermentation performance under high-gravity conditions. Appl. Environ. Microbiol. 73: 815-824

7. González SS, Gallo L, Climent MA, Barrio E, Querol A. et al. (2007) Enological characterization of natural hybrids from *Saccharomyces cerevisiae* and *S. kudriavzevii*. Int. J. Food Microbiol 116, 11-17.

8. Tirosh I, Reikhav S, Levy AA, Barkai N. (2009) A yeast hybrid provides insight into the evolution of gene expression regulation. Science 324: 659-662.

9. Borneman AR**,** Gianoulis TA, Zhang ZD, Yu H, Rozowsky J, et al. (2007) Divergence of transcription factor binding sites across related yeast species. Science 317: 815-819**.**

10. Pu S, Wong J, Turner B, Cho E, Wodak SJ (2009) Up-to-date catalogues of yeast protein complexes. Nucleic Acids Res **37**: 825–831.

11. Gavin AC, Bösche M, Krause R, Grandi P, Marzioch M, et al. (2002) Functional organization of the yeast proteome by systematic analysis of protein complexes. Nature 415: 141-147.

12. Tarassov K, Messier V, Landry CR, Radinovic S, Serna Molina MM, et al. (2008) An in vivo map of the yeast protein interactome. Science 320: 1465-1470.

13. Steel GJ, Brownsword J, Stirling CJ. (2002) Tail-anchored protein insertion into yeast ER requires a novel posttranslational mechanism which is independent of the SEC machinery. Biochemistry 41: 11914-11920

14. Prasad R, Niederberger P, Hütter R (1987) Tryptophan accumulation in *Saccharomyces cerevisiae* under the influence of an artificial yeast TRP gene cluster. Yeast 3: 95-105

15. Cho EJ, Kobor MS, Kim M, Greenblatt J, Buratowski S. (2001) Opposing effects of Ctk1 kinase and Fep1 phosphatase at Ser 2 of the RNA polymerase II C- terminal domain. Genes Dev 15: 3319-3329

16. Bean JM, Siggia ED, Cross FR. (2005) High functional overlap between Mlu I cell- cycle box binding factor in the G1/S transcriptional program in *Saccharomyces cerevisiae*. Genetics 171: 49-61.

17. He B, Chen P, Chen SY, Vancura KL, Michaelis S, et al. (1991) *RAM2*, an essential gene of yeast, and *RAM1* encode the two polypeptide components of the farnesyltransferase that prenylates a-factor and Ras proteins. Proc Natl Acad Sci U S A 88: 11373-11377.

18. Tam ATY, Pike BL, Hammet A, Heierhorst J. (2007) Telomere-related function of yeast KU in the repair of bleomycin-induced DNA damage. Biochem Biophys Res Commun 357: 800-803

19. Liti G, Haricharan S, Cubillos FA, Tierney AL, Sharp S, *et al.* (2009) Segregating *YKU80* and *TLC1* alleles underlying natural variation in telomere properties in wild yeast. *PLoS Genet.* **5**: e1000659.

20. Steinmetz LM, Scharfe C, Deutschbauer AM, Mokranjac D, Herman ZS, *et al.* (2002) Systematic screen for human disease genes in yeast. Nat Genet 31: 400-404

21. Lang, G. I. Murray AW, Botstein D. (2009) The cost of gene expression underlies a fitness trade-off in yeast. Proc Natl Acad Sci U S A 106: 5755-5760.

22. Ohno S. (1970) Evolution by gene duplication. New York Springer-Verlag.

23. Papp B, Pál C, Hurst LD (2003) Dosage sensitivity and the evolution of gene families in yeast. Nature 424: 194-197.

24. Isalan M, Lemerle C, Michalodimitrakis K, Horn C, Beltrao P et al. (2008) Evolvability and hierarchy in rewired bacterial gene networks. Nature 452: 840-845.

25. Delneri D, Colson I, Grammenoudi S, Roberts IN, Louis EJ, *et al.* (2003) Engineering evolution to study speciation in yeasts. Nature 422: 68-72

26. Rigaut G, Shevchenko A, Rutz B, Wilm M, Mann M, *et al.* (1999) A generic protein purification method for protein complex characterization and proteome exploration. Nat Biotechnol 17: 1030-1032

27. Shevchenko A, Tomas H, Havlis J, Olsen JV, Mann M. (2006) In-gel digestion for mass spectrometric characterization of proteins and proteomes. Nat Protoc 1: 2856-2860

28. Nesvizhskii AI, Keller A, Kolker E, Aebersold RA (2003) Statistical model for identifying proteins by tandem mass spectrometry. Anal Chem 75: 4646-4658.

29. Carter Z, Delneri D. (2010) New generation of loxP-mutated deletion cassettes for the genetic manipulation of yeast natural isolates. Yeast 27: 765-775.

30. Delneri, D. (2011) Competition experiments coupled with high-throughput analyses for functional genomics studies in yeast. Methods Mol Biol 759: 271-282.

**FIGURE LEGENDS**

**Figure 1**: TAP-strategy for recovery and identification of hybrid protein complexes. *S. cerevisiae* strains with the TAP cassette inserted into the C-terminal of one member of the complex (TAP-tag A) were crossed with *S. mikatae* and *S. bayanus* species. The complexes that freely formed in the hybrids were then isolated and the interacting members identified via MS analysis. A', B' and C' represent the homologs of the *S. cerevisiae* A, B, C proteins, respectively.

**Figure 2**: Peptide map of the *S. bayanus* Sec63p. The peptides in common for both *S. cerevisiae* and *S. bayanus* species are shown as green boxes, while *S. bayanus* specific peptides are shown as pink boxes. Unique peptides detected independently in different biological repeats are marked with asterisks. The MS spectra of unique *S. bayanus* ion peptides T42 and T53 are shown below.

**Figure 3**: Fitness assays of *Sc/Sb* hybrids carrying different type of *TRP2/TRP3* chimeric complexes. *Sc/Sb* hybrids were genetically modified to carry either the two different chimeric complexes, Trp2p$^{Sb}$/Trp3p$^{Sc}$ and Trp2p$^{Sc}$/Trp3p$^{Sb}$, or the two parental hemizygous controls, Trp2p$^{Sb}$/Trp3p$^{Sb}$ and Trp2p$^{Sc}$/Trp3p$^{Sc}$ (panel A). The growth curves of the engineered hybrids shows that Trp2p$^{Sb}$/Trp3p$^{Sc}$ grows better than the other combinations in F1 minimal media lacking tryptophan (panel B). Fitness competition essay between *Sc/Sb* hybrids, carrying different combination of the *TRP2/TRP3* complex, and the GFP reference strain shows again that Trp2p$^{Sb}$/Trp3p$^{Sc}$ grows faster (panel C).

**Figure 4**: Growth essays of *Sc/Sb* hybrids carrying different types of MBF chimeric complexes. *Sc/Sb* hybrids were genetically modified either to carry the two different chimeric complexes, Mbp1$^{Sb}$/Swi6$^{Sc}$ and Mbp1$^{Sc}$/Swi6$^{Sb}$, or the two uni-parental controls, Mbp1$^{Sb}$/Swi6$^{Sb}$ and Mbp1$^{Sc}$/Swi6$^{Sc}$ (Panel A). The growth spot essay of the engineered hybrids in rich YPD and YP-glycerol media are shown in Panel B. The strain carrying the S. bayanus homologous Mbp1$^{Sb}$ and Swi6$^{Sb}$ is the only one that performs respiratory growth and grow normally in the presence of glycerol a sole carbon source.

Figure 1

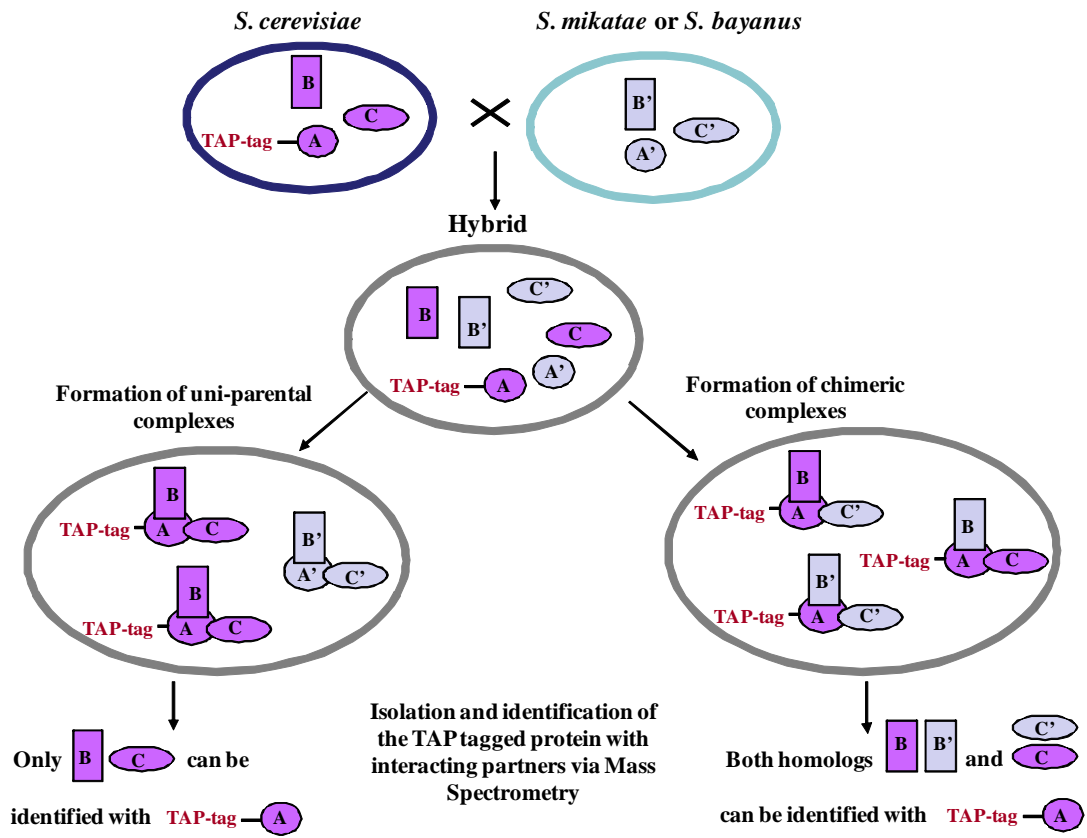

Figure 2

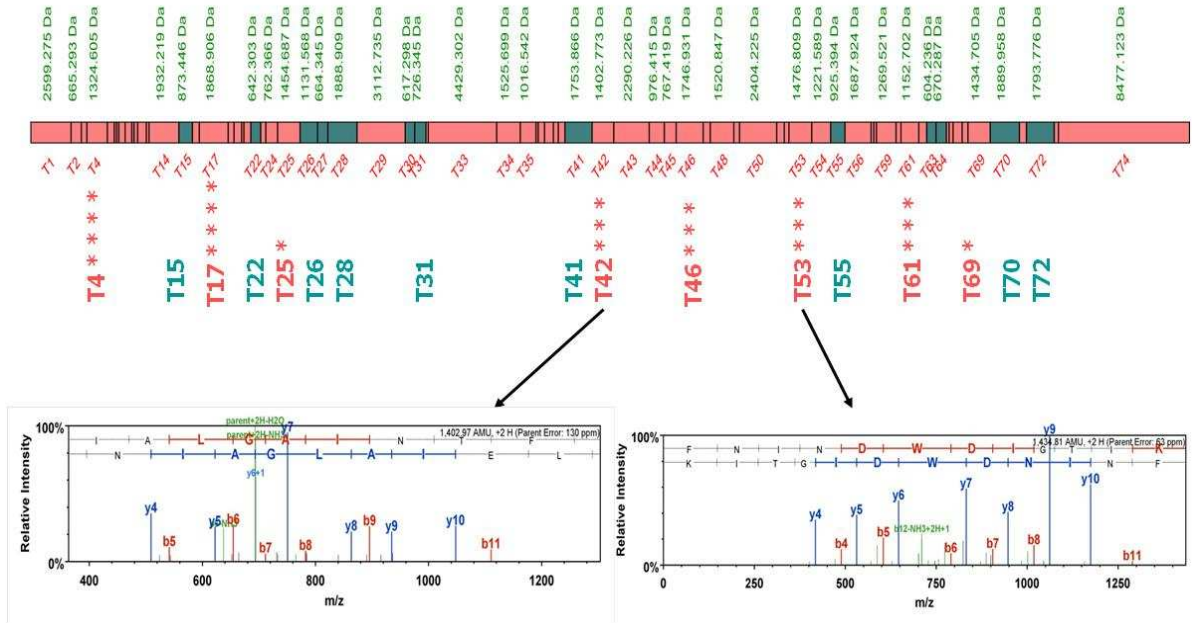

Figure 3

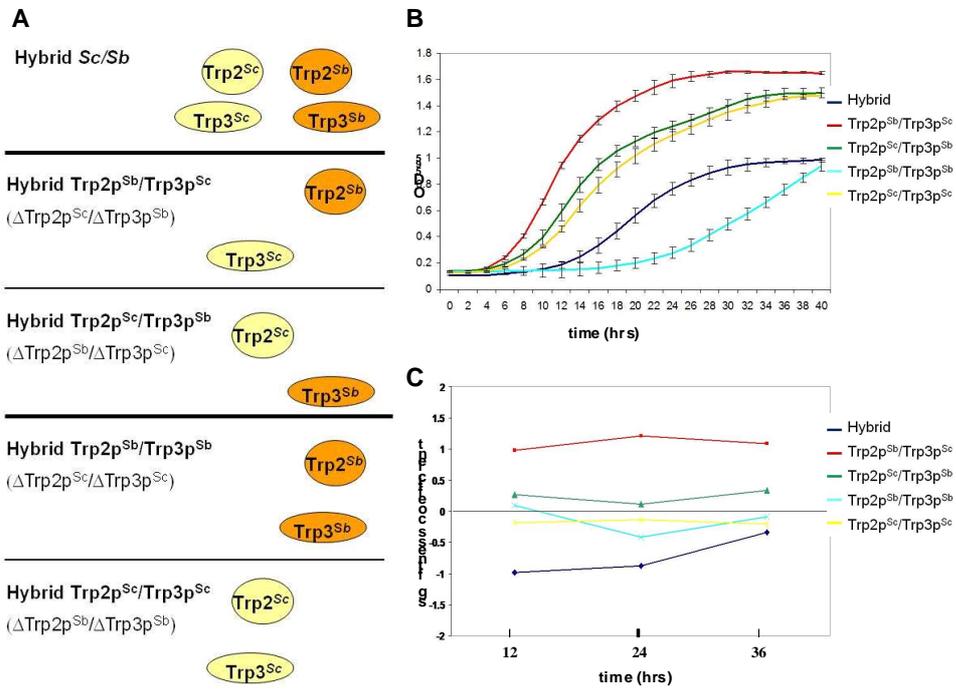

Figure 4

A

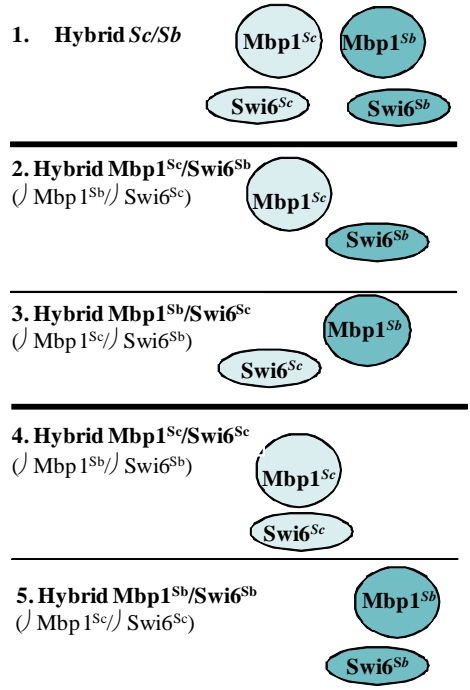

1. Hybrid *Sc/Sb*

2. Hybrid Mbp1$^{Sc}$/Swi6$^{Sb}$
   (∆ Mbp1$^{Sb}$/∆ Swi6$^{Sc}$)

3. Hybrid Mbp1$^{Sb}$/Swi6$^{Sc}$
   (∆ Mbp1$^{Sc}$/∆ Swi6$^{Sb}$)

4. Hybrid Mbp1$^{Sc}$/Swi6$^{Sc}$
   (∆ Mbp1$^{Sb}$/∆ Swi6$^{Sb}$)

5. Hybrid Mbp1$^{Sb}$/Swi6$^{Sb}$
   (∆ Mbp1$^{Sc}$/∆ Swi6$^{Sc}$)

B

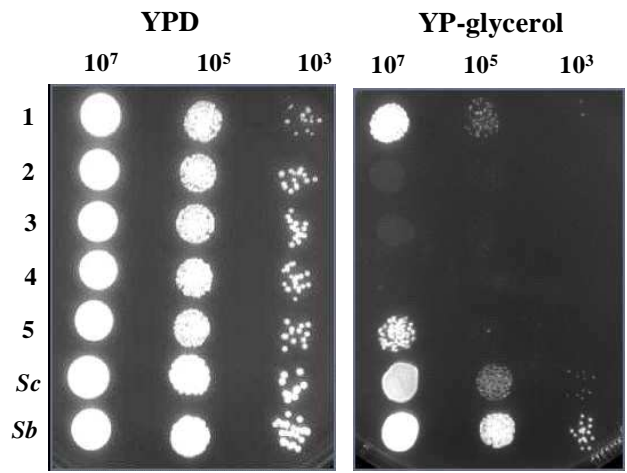